\documentclass[aps,prb,reprint]{revtex4-1}
\pdfoutput=1

\usepackage[pdftex]{graphicx} 
\usepackage{amsmath}
\usepackage{amsfonts}
\usepackage{mathrsfs}
\usepackage{bm}

\usepackage{times}
 
\begin{document}

\title{Fulde-Ferrell--Larkin-Ovchinnikov state in the dimensional crossover between one- and three-dimensional lattices}
\author{Dong-Hee Kim}
\affiliation{Department of Applied Physics and Centre of Excellence in Computational Nanoscience (COMP), Aalto University, FI-00076 Aalto, Finland}
\author{P\"{a}ivi T\"{o}rm\"{a}}
\email{paivi.torma@aalto.fi}
\affiliation{Department of Applied Physics and Centre of Excellence in Computational Nanoscience (COMP), Aalto University, FI-00076 Aalto, Finland}

\begin{abstract}
We present a full phase diagram for the one-dimensional (1D) to three-dimensional (3D) crossover of 
the Fulde-Ferrell--Larkin-Ovchinnikov (FFLO) state 
in an attractive Hubbard model of 3D-coupled chains in a harmonic trap. 
We employ real-space dynamical mean-field theory which describes full
local quantum fluctuations beyond the usual mean-field and local density approximation.
We find strong dimensionality effects on the shell structure undergoing 
a crossover between distinctive quasi-1D and quasi-3D regimes.
We predict an optimal regime for the FFLO state that is considerably extended to 
intermediate interchain couplings and polarizations, 
directly realizable with ultracold atomic gases. 
We find that the 1D-like FFLO feature is vulnerable to thermal fluctuations, 
while the FFLO state of mixed 1D-3D character can be stabilized at a higher temperature.
\end{abstract}

\maketitle

The interplay between fermion pairing and magnetism is at the heart of 
understanding strongly correlated systems ranging from unconventional superconductors
and ultracold gases to neutron stars and quarks.~\cite{Casalbuoni2004}
BCS-type superconductivity is suppressed by a large magnetic field 
exceeding the Chandrasekhar--Clogston limit. However, it has been proposed 
as a paradigm of superconductivity in high magnetic fields 
that it is possible for superconductivity and magnetism to coexist with 
exotic pairing mechanisms.~\cite{FF,LO,Sarma1963,Liu2003}
The Fulde-Ferrell--Larkin-Ovchinnikov (FFLO) state would arise
with this interplay,~\cite{FF,LO} but it still remains elusive 
in spite of indirect experimental evidence observed.~\cite{Radovan2003}
The FFLO state is characterized 
by the Cooper pair carrying finite momentum causing a spatially modulated 
order parameter. 
One peculiar feature of this exotic phase is that apparently
its stability is largely affected by the dimensionality of the system.
It turns out that the three-dimensional (3D)-FFLO state occupies a thin area of 
the mean-field phase diagram~\cite{Sheehy2006}
though the signature can be stronger in the systems that support nesting
such as optical lattices~\cite{Koponen2007,Loh2010} and elongated traps.~\cite{Kim2011}
Indeed, only phase separation was observed for ultracold gases in 3D traps.~\cite{Zwierlein2006,Partridge2006,Nascimbene2009}
On the other hand, in the exact solution of a one-dimensional (1D) system, the FFLO character appears 
at any finite spin-polarization,~\cite{Yang2001} while long-range order cannot exist in 1D.
The experiment~\cite{Liao2010} in 1D was consistent with the FFLO theory, 
although the state remains unidentified.

A natural question arising is whether one can combine the promising 1D-FFLO features 
and long-range order provided by higher dimensions. 
This has been considered in previous mean-field~\cite{Parish2007} and effective field theory~\cite{Zhao2008} 
studies for coupled continuum-1D gases, and for the Hubbard ladder.~\cite{Feiguin2009}
The trapping potential is essential in ultracold gas experiments, and thus 
in spin-polarized systems, one can expect a shell structure of different phases along the trap. 
Therefore, beyond the local density approximation, the inhomogeneous superfluid 
and normal phases need to be treated in a unified framework by including full local quantum fluctuations.
Neglecting local quantum fluctuations creates an apparent bias 
in favor of the superfluid state and against the normal state.~\cite{Kim2011}
Here, using a real-space dynamical mean-field theory (DMFT), we investigate the 1D-3D crossover problem 
within the anisotropic Hubbard model in a trap.

The dimensionality effect to the FFLO state that we are considering here
differs from the two-dimensional (2D)-3D crossover studied in layered superconductors.~\cite{Matsuda2007}
There, the quasi-2D character minimizes the orbital pair breaking effects 
in a magnetic field, and the Zeeman effect may then lead to a $d$-wave FFLO state. 
Here we consider $s$-wave pairing, and orbital effects are absent. 
Note that a particle-hole transformation can bring in another interesting perspective 
by mapping the FFLO state to the striped phase of the doped 
repulsive Hubbard model.~\cite{Moreo2007}
This emphasizes the importance of the FFLO state in the general context of 
high-$T_\mathrm{c}$ superconductivity. 

We perform a real-space variant of DMFT calculations 
on the attractive Hubbard model of 3D-coupled chains,
\begin{eqnarray*}
\mathcal{H} = && -t_\parallel \sum_{il\sigma} 
(c^\dagger_{il\sigma}c_{(i+1)l\sigma} + \mathrm{h.c.}) 
-t_\perp \sum_{\langle l l^\prime \rangle} \sum_{i\sigma}
c^\dagger_{il\sigma}c_{il^\prime\sigma} \\
&& + U \sum_{il} \hat{n}_{il\uparrow} \hat{n}_{il\downarrow} 
+ \sum_{il\sigma} (V_i - \mu_\sigma) \hat{n}_{il\sigma},
\end{eqnarray*}
where $c^\dagger_{il\sigma}$($c_{il\sigma}$) creates (annihilates) 
a fermion with spin $\sigma=\uparrow,\downarrow$ at site $i$ of chain $l$, the density 
$\hat{n}_{il\sigma} = c^\dagger_{il\sigma} c_{il\sigma}$. We define the superfluid 
order parameter as $\Delta=-\langle c^\dagger_{\uparrow} c^\dagger_{\downarrow}\rangle$.
Throughout the calculations, the hopping $t_\parallel$ is set to unity. 
The dimensionality is thus tuned by the interchain coupling $t_\perp$ and varied
from 1D ($t_\perp=0$) to 3D ($t_\perp=1$). The chemical potentials $\mu_\uparrow$ 
and $\mu_\downarrow$ control the polarization 
$P=(N_\uparrow-N_\downarrow)/(N_\uparrow+N_\downarrow)$ by keeping
the total particle number $N_\uparrow+N_\downarrow \sim 120$. 
The harmonic potential trapping the gases in a longitudinal chain is 
given as $V_i=5\times 10^{-5}(i-1/2)^2$.
The on-site interaction $U$ is selected to be the value corresponding to unitarity 
where the 3D two-body scattering length diverges.~\cite{Burovski2006} 
The value of $U$ varies from $-2.038$ ($t_\perp=0.1$) to $-7.915$ ($t_\perp=1$), 
which is comparable to the half bandwidths. 

In order to treat an inhomogeneous phase along the chain, 
a site-dependent self-energy $\bm{\Sigma}_i$ is considered 
within DMFT.~\cite{DMFTreview,Snoek2011,Kim2011}
With translational invariance in transverse directions, 
the on-chain Green's function is written for sites $i,j$ and 
transverse momentum $\bm{k}_\perp$ as
\begin{equation*}
[\bm{G}^{-1}(\bm{k}_\perp;i\omega_n)]_{ij} =
[i\omega_n \bm{\sigma}_0- \epsilon_{\bm{k}_\perp} \bm{\sigma}_3 
- \bm{\Sigma}_{i}(i\omega_n) ] \delta_{ij} -\bm{h}^\parallel_{ij},
\end{equation*}
where $\omega_n$ denotes the Matsubara frequency, 
$\bm{\sigma}$ is the Pauli matrix,
$\bm{h}^\parallel$ is the non-interacting part of the chain Hamiltonian,
the 2D energy dispersion $\epsilon_{\bm{k}_\perp} \equiv -2t_\perp (\cos k_x + \cos k_y)$ 
with $\bm{k}_\perp=(k_x,k_y)$, and the operators are in the Nambu basis.
In the self-consistency of DMFT, the impurity Green's function is obtained as  
$\bm{\mathcal{G}}_{i}(i\omega_n) = \sum_{\bm{k}_\perp} \bm{G}_{ii}(\bm{k}_\perp;i\omega_n)$.
While we have a local but site-dependent self-energy term, the procedures 
can be compared to the chain-DMFT.~\cite{Biermann2001}
The exact diagonalization method is employed to solve 
the site-dependent impurity problem.~\cite{Kim2011}

With the on-site interaction $U=-4$, we have found that 
our calculations qualitatively reproduce the two important characteristics 
of strongly interacting 1D Fermi gases.~\cite{Yang2001,Orso2007}
First, at small polarizations,
the trap edges are fully paired while the trap center is partially polarized. 
The entire area becomes polarized as the total polarization increases. 
Second, at finite polarizations, the polarized trap center is associated 
with the FFLO state exhibiting a spatially oscillating order parameter. 

Clear features of getting away from the 1D limit are observed. 
For the studied finite interchain couplings, the pairing at the edges, 
namely the first 1D feature listed above, is easily broken at small polarizations.
This indicates that immediately away from 1D, one can observe fully polarized edges 
much before the whole area gets polarized. In contrast, it turns out that 
the other part of the 1D features, the emergence of a partially polarized center 
in fully paired vicinities, survives at small interchain couplings. 
Away from but close to the 1D limit, we typically find
a shell structure of polarized edges, fully paired shoulders, and a partially polarized center.

\begin{figure}
\includegraphics[width=0.48\textwidth]{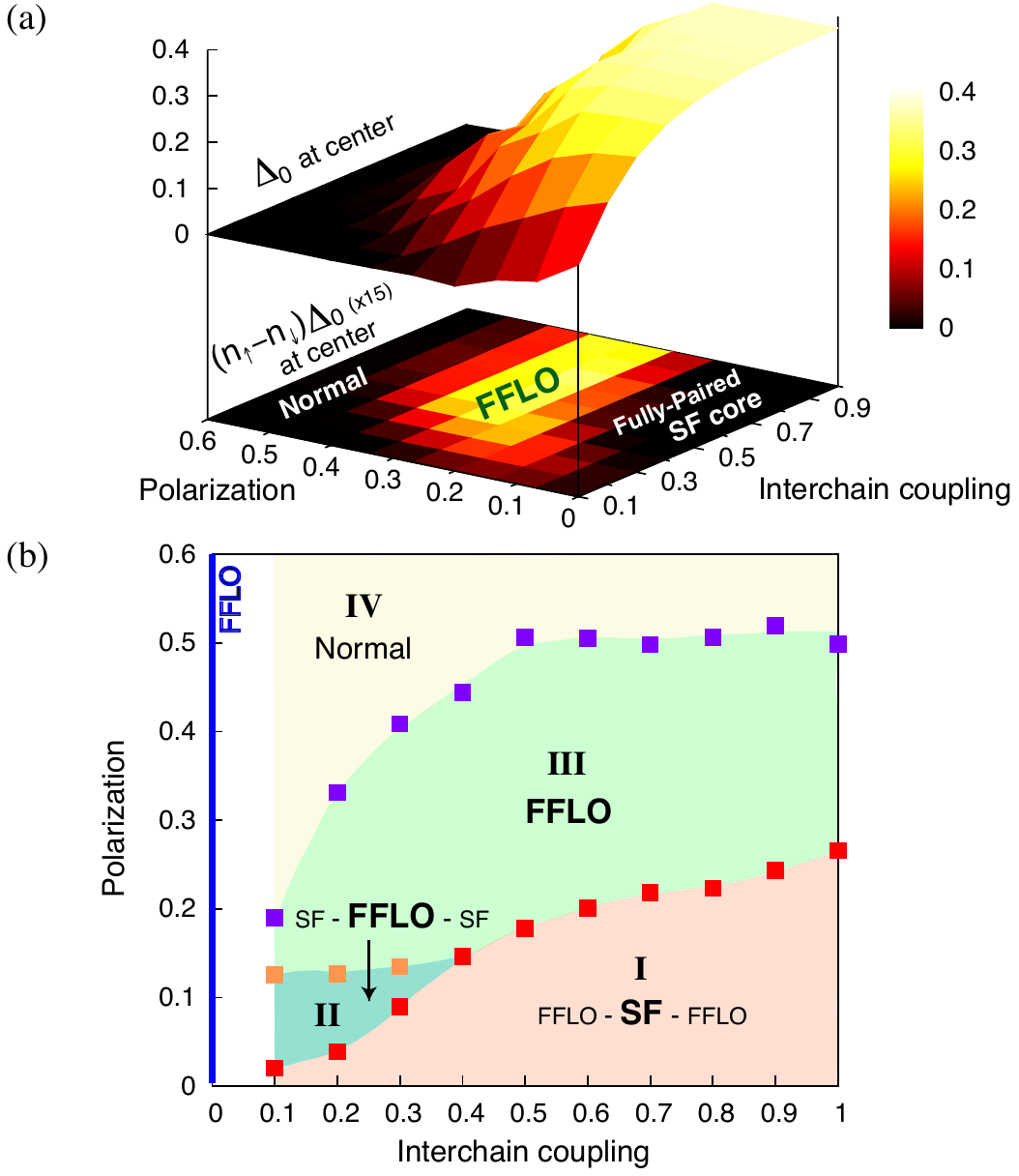}
\caption{\label{fig1}
(Color online) Phase diagram of the 3D coupled-chain Hubbard model.
The particle densities $n_{\uparrow,\downarrow}$ and the order parameter $\Delta$ 
are calculated as a function of the interchain coupling and the polarization at zero temperature.
(a) The oscillation amplitude $\Delta_0$ of the order parameter and 
the density difference $n_\uparrow - n_\downarrow$ at the trap center. 
The phase at the trap center can be divided into three: the fully-paired superfluid (SF)
($n_\uparrow =n_\downarrow, \Delta\neq 0$), FFLO (oscillating $n$ and $\Delta$), 
and normal ($\Delta = 0$) phases.
(b) Phases I--III associated with the shell structures in the trap, explained in the text. 
}
\end{figure}

In Fig.~\ref{fig1}, we present a real-space DMFT phase diagram of the Hubbard model 
of 3D-coupled chains at zero temperature.  
Considering only the phase at the trap center, we find that the emergence of 
the FFLO-type oscillating order parameter occupies a large area 
of polarizations and interchain couplings. This wide coexistence area of
a finite density difference and a finite order parameter 
extends all the way to the 3D limit, as opposed to the mean-field phase diagram 
on a 3D continuum where the FFLO phase occupies only a tiny area.~\cite{Sheehy2006}

Our zero-temperature phase diagram characterizes the emergence of three types of 
shell structures, as shown in Fig.~\ref{fig1}(b).
First, area I shows quasi-3D features in the shell structure where the fully paired 
superfluid (SF) core exists. There are FFLO-type oscillations found between the SF 
core and the fully polarized edges.
Second, area II shows an inverted sequence: an FFLO core surrounded 
by fully paired shoulders. The edges are polarized in this area, 
and small order parameter oscillations also exist at the interfaces 
between the fully paired shoulders and the fully polarized edges. 
While this indicates a mixture of 1D and 3D features,
area II can be identified as a quasi-1D phase because of the 
spatial pattern of the FFLO oscillations emerging at the trap center. 
Area II is only found at interchain couplings $t_\perp \le0.3$.
Third, area III has a two-shell structure where
the FFLO-type oscillations reside in the entire area 
of the partially polarized core, surrounded by fully polarized edges. 
Area III is found across the whole range of interchain couplings at intermediate-high
polarizations below the transition to the normal phase.

\begin{figure*}[t]
\includegraphics[width=0.98\textwidth]{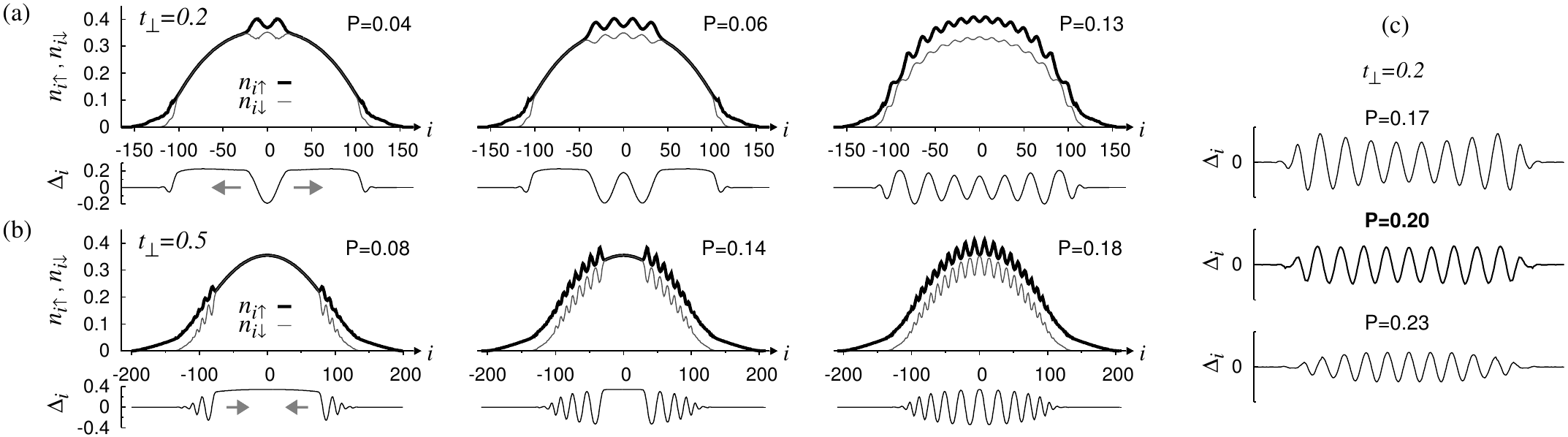}
\caption{\label{fig2}
Evolution of the FFLO oscillations in the 1D-3D crossover. 
The profiles of the particle densities $n_{\uparrow, \downarrow}$ 
and the order parameter $\Delta$ along the chain sites $i$ are presented 
for the two regimes of interchain couplings (a) $t_\perp=0.2$ (quasi-1D) and (b) $t_\perp=0.5$ (quasi-3D).
In the quasi-1D regime, for small polarization $P=0.04$, the FFLO-type oscillations 
at the center are surrounded by the fully paired shoulders.
The region of oscillating $\Delta$ develops from the center, expands
toward the edges as $P$ increases, and then emerges over 
the whole area for $P=0.13$.
On the contrary, in the quasi-3D regime, the oscillations are initially at the partially polarized 
intermediate regions and spread toward the center as $P$ increases.
At finite interchain couplings, the far-edges of the trap are always polarized.
In (c), the evolution of the oscillation envelope of $\Delta$ is shown with increasing $P$ at $t_\perp=0.2$
in the quasi-1D regime.
Uniform oscillations occur along the trap at an intermediate polarization.   
}
\end{figure*}

We find that the FFLO character evolves very differently with polarization 
in the quasi-1D and quasi-3D regimes. Figure~\ref{fig2} shows the shell structures 
with increasing polarization $P$ at both sides of the 1D-3D crossover. 
In the quasi-1D regime [Fig.~\ref{fig2}(a)], the evolution is dominated by the expansion of the
FFLO core. In contrast, in the quasi-3D regime [Fig.~\ref{fig2}(b)], the fully paired core shrinks 
with increasing $P$ while the FFLO-type oscillations at the shoulders
move toward the trap center. 
Near the crossover, these quasi-1D and quasi-3D features coexist. 
At the interchain coupling $t_\perp=0.3$, the oscillations of the order parameter $\Delta$
become significant at both the trap center and edges, and the fully paired shoulders 
decrease from both sides as $P$ increases.
In addition, when $P$ is increased to enter the area III, 
the oscillations of $\Delta$ along the trap show
a distinct feature: At small $t_\perp$, the amplitude of $\Delta$ 
becomes spatially uniform, despite the trap, at intermediate $P$ [see Fig.~\ref{fig2}(c)]. 
This can be compared to the zero derivative of the FFLO momentum
as a function of the chemical potential.~\cite{Parish2007}
The uniform oscillations occur only in the quasi-1D regime, raising the possibility
that a peak signal of the FFLO momentum is more visible in this regime. 

Our phase diagram suggests that the optimum spot for observing the FFLO state 
is extended over a significantly large area of the dimensional crossover. 
In particular, the largest range of polarizations associated with the FFLO-featured areas (II+III) 
at the trap center is found around interchain coupling $t_\perp \sim 0.4$. 
Since we still have two control parameters, the onsite interaction and the temperature, 
the phase diagram should be further generalized to discuss ultracold gas experiments 
which are conducted at finite temperatures and with a tunable atomic scattering length.
We have examined also a stronger interaction $U=-4$ at small $t_\perp$'s, including 
the 1D limit ($t_\perp=0$) where the 3D scattering approach fails. 
At small interchain couplings, we have found that the main effect of a stronger interaction 
is to increase the critical polarization of the transition to the normal phase while the change in 
the phase II area is relatively small. 
In the 3D limit, stronger interactions tend to keep the fully paired SF core even 
at larger polarizations, as observed in the previous experiments in the BEC regime,~\cite{Zwierlein2006}
which can make the 3D-FFLO area much smaller.

\begin{figure}[b]
\includegraphics[width=0.48\textwidth]{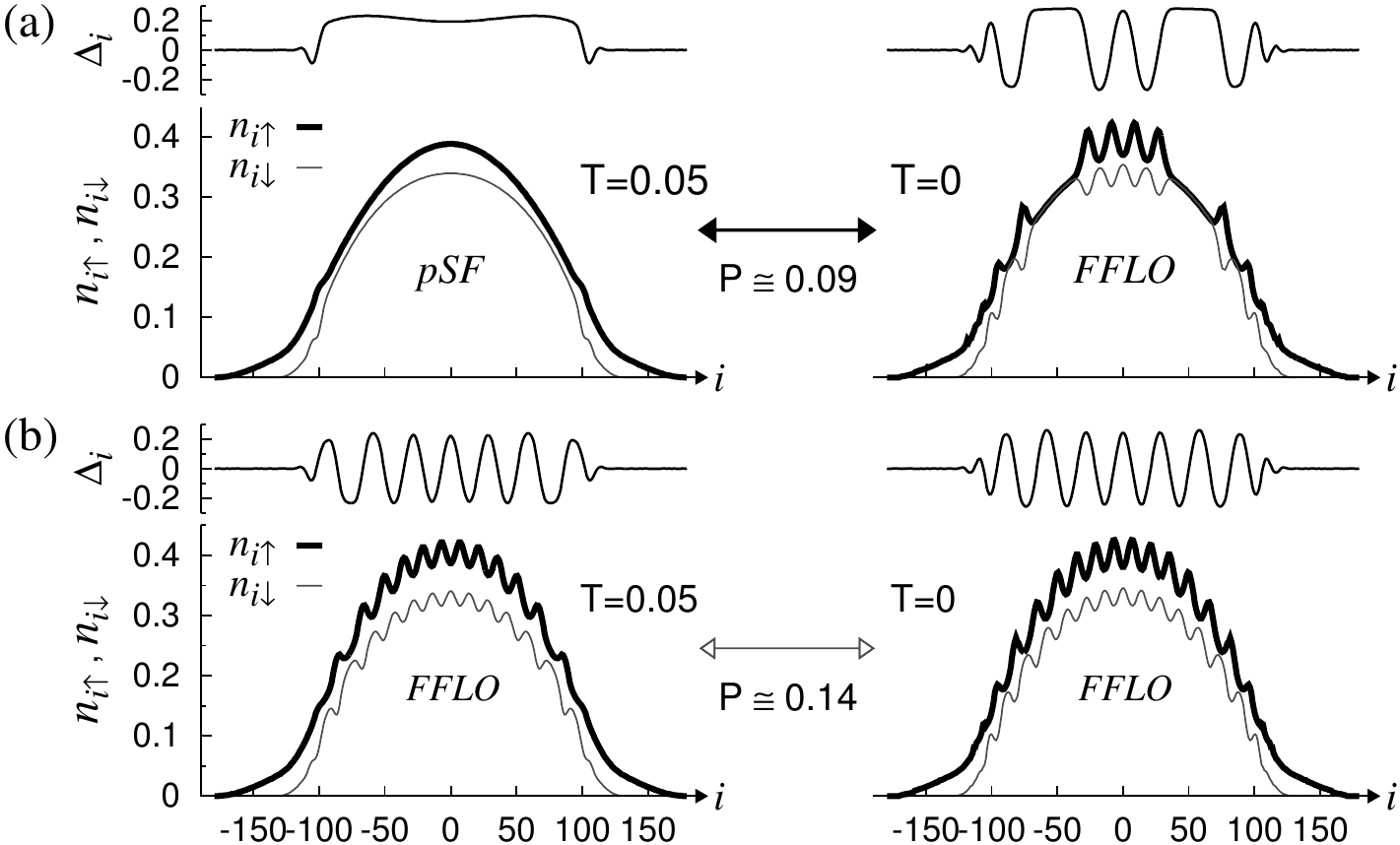}
\caption{\label{fig3}
Polarized superfluid phase at finite temperature. 
For interchain coupling $t_\perp=0.3$, the profiles at temperature $T=0.05$ 
are compared with those at $T=0$ for  the particle densities $n_{\uparrow,\downarrow}$ 
and the order parameter $\Delta$ along the chain sites $i$.
At polarization (a) $P \simeq 0.09$, corresponding to phase II, the structure of 
the FFLO core surrounded by fully paired shoulders 
found at $T=0$ is completely changed into the polarized superfluid phase with a uniform 
order parameter at $T=0.05$. The order parameter oscillations at the edges are 
still observed at $T=0.05$.
In contrast, at a higher polarization (b) $P\simeq 0.14$, corresponding to phase III,
similar FFLO characteristics are identified at both temperatures $T=0.05$ and $T=0$. 
}
\end{figure}

The stability of the FFLO state at finite temperature is a crucial question.
We have examined the behavior at a finite temperature $T=0.05$ 
near the dimensional crossover at the interchain coupling $t_\perp=0.3$. 
A low temperature algorithm is used.~\cite{Capone2007}
We find that finite temperature can stabilize the polarized superfluid (pSF) state 
at small polarizations while it destroys the quasi-1D FFLO character.
Figure~\ref{fig3} shows comparisons between the density and order parameter profiles
at two different temperatures $T=0.05$ and $T=0$. 
These profiles indicate that area II with the FFLO core
melts at $T=0.05$ into the pSF state, similar to the results in 1D.~\cite{Wolak2010}
In the case of shell structures, the polarization from the FFLO area is easily redistributed 
in the trap at finite temperature to create a BCS-type order parameter
with extra majority particles accommodated as thermal quasiparticles.
In contrast, it turns out that area III with a trap-wide FFLO character 
at higher polarizations is not affected by the finite temperature examined. 

A comparison with the previous mean-field results in 1D and 3D lattices shows
the drastic effects of local quantum fluctuations: 
(1) Our results indicate that the large size of the FFLO area predicted 
by mean-field theory in 3D lattices~\cite{Koponen2007,Loh2010} 
may have been overestimated.
In the 3D limit, we find that the FFLO state is broken near local polarization 
$0.35$ at a trap center that is approximately a quarter filled,
which contrasts to the mean-field values
$\sim 0.6$ ($|U|=5.14$)~\cite{Koponen2007} 
and $\sim0.75$ ($|U|=6$)~\cite{Loh2010} at the same filling.
This enhancement of the normal state can be understood in light of the fact that
including the particle-hole channel was shown to reduce pairing significantly in lattices:~\cite{Kim2009}
DMFT includes full local quantum fluctuations causing such higher order effects. 
(2) In 1D lattices, mean-field theory \cite{Koponen2008} predicted only a 3D-like shell 
structure with a fully paired center and polarized edges.
In contrast, we find the reversal of the shell structure, 
similar to what has been predicted by continuum mean-field study.~\cite{Parish2007}
(3) In addition, our characterization of the FFLO state does not find 
the well-separated domain wall between the sign changes 
of the order parameter predicted by the mean-field theories in lattices~\cite{Loh2010} 
and in continuum.~\cite{Parish2007}
Our findings, the drastic decrease of critical polarization, the shell structure reversal,
and the absence of domain walls,
emphasize the importance of local quantum fluctuations in lattices, regardless of the dimensionality.

Our findings are directly applicable to future experiments. 
Recent experiments realized a weakly coupled 2D array of 1D tubes with ultracold $^6$Li 
gases, observing the density profile characteristics of attractively interacting spin-polarized 
1D Fermi gases.~\cite{Liao2010} This system can be further extended toward
the quasi-1D regime by adjusting the optical lattice potentials, and if needed, 
discreteness along the tube direction can be realized by 3D lattices.~\cite{Jordens2008,Schneider2008}
Although it is nontrivial to experimentally reveal the oscillating order parameter, 
several methods have been recently suggested to detect the FFLO state 
particularly in 1D systems, see~\cite{Edge2009,Kajala2011} and references therein.
Our DMFT calculations have shown that the FFLO phase occupies 
a significant area of the phase diagram throughout the 1D-3D crossover.
In addition, our calculations at finite temperature have found that the FFLO character is 
more stable at intermediate rather than small polarizations where the phase becomes 
the polarized superfluid at low temperature. 
Our findings will help identify the presence of the FFLO state in ultracold atomic gases.  
In a general perspective,  by using beyond-mean-field calculations, we have investigated
the dimensionality effects on the existence of the FFLO state in 1D-3D lattice systems.
This may potentially help to understand also the puzzling repulsive Hubbard model counterparts 
such as the striped phase. 

\begin{acknowledgments}
This work was supported by the Academy of Finland through 
its Centers of Excellence Programme (2012-2017) 
and under Projects No. 139514, No. 141039, No. 213362, and No. 135000 
and conducted as a part of a EURYI scheme grant (see www.esf.org/euryi). 
Computing resources were provided by CSC--the Finnish IT Centre for Science 
and the Triton cluster at Aalto University. 
\end{acknowledgments}


\begin{thebibliography}{}

\bibitem{Casalbuoni2004}
R. Casalbuoni and G. Nardulli, 
Rev. Mod. Phys. {\bf 76}, 263 (2004). 

\bibitem{FF}
P. Fulde and R. A. Ferrell, 
Phys. Rev. {\bf 135}, A550 (1964).

\bibitem{LO}
A. I. Larkin and Y. N. Ovchinnikov, 
Zh. Eksp. Teor. Fiz. {\bf 47}, 1136 (1964) [Sov. Phys. JETP {\bf 20}, 762 (1965)]. 

\bibitem{Sarma1963}
G. Sarma, 
J. Phys. Chem. Solids {\bf 24}, 1029 (1963).

\bibitem{Liu2003}
W. V. Liu and F. Wilczek, 
Phys. Rev. Lett {\bf 90}, 047002 (2003).

\bibitem{Radovan2003}
H. A. Radovan, N. A. Fortune, T. P. Murphy, S. T. Hannahs, E. C. Palm, S. W. Tozer, and D. Hall, 
Nature (London) {\bf 425}, 51 (2003).

\bibitem{Sheehy2006}
D. E. Sheehy and L. Radzihovsky, 
Phys. Rev. Lett. {\bf 96}, 060401 (2006).

\bibitem{Koponen2007}
T. K. Koponen, T. Paananen, J.-P. Martikainen, and P. T\"orm\"a,
Phys. Rev. Lett. {\bf 99}, 120403 (2007).

\bibitem{Loh2010}
Y. L. Loh and N. Trivedi,
Phys. Rev. Lett. {\bf 104}, 165302 (2010).

\bibitem{Kim2011}
D.-H. Kim, J. J. Kinnunen, J.-P. Martikainen, and P. T\"orm\"a,
Phys. Rev. Lett. {\bf 106}, 095301 (2011).

\bibitem{Zwierlein2006}
M. W. Zwierlein, A. Schirotzek, C. H. Schunck, and W. Ketterle,
Science {\bf 311}, 492 (2006).

\bibitem{Partridge2006}
G. B. Partridge, W. Li, R. I. Kamar, Y. Liao, and R. G. Hulet,
Science {\bf 311}, 503 (2006).

\bibitem{Nascimbene2009}
S. Nascimb\'ene, N. Navon, K. J. Jiang, L. Tarruell, M. Teichmann, J. McKeever, F. Chevy, and C. Salomon,
Phys. Rev. Lett. {\bf 103}, 170402 (2009).

\bibitem{Yang2001}
K. Yang, 
Phys. Rev. B {\bf 63}, 140511(R) (2001).

\bibitem{Liao2010}
Y. Liao, A. S. C. Rittner, T. Paprotta, W. Li, G. B. Partridge, R. G. Hulet, S. K. Baur, and E. J. Mueller,
Nature (London) {\bf 467}, 567 (2010).

\bibitem{Parish2007}
M. M. Parish, S. K. Baur, E. J. Mueller, and D. A. Huse,
Phys. Rev. Lett. {\bf 99}, 250403 (2007).

\bibitem{Zhao2008}
E. Zhao and W. V. Liu,
Phys. Rev. A {\bf 78}, 063605 (2008).

\bibitem{Feiguin2009}
A. E. Feiguin and F. Heidrich-Meisner,
Phys. Rev. Lett. {\bf 102}, 076403 (2009).

\bibitem{Matsuda2007}
Y. Matsuda and H. Shimahara,
J. Phys. Soc. Jpn. {\bf 76}, 051005 (2007).

\bibitem{Moreo2007}
A. Moreo and D. J. Scalapino,
Phys. Rev. Lett. {\bf 98}, 216402 (2007).

\bibitem{Burovski2006}
E. Burovski, N. Prokof'ev, B. Svistunov, and M. Troyer,
Phys. Rev. Lett. {\bf 96}, 160402 (2006).

\bibitem{DMFTreview}
A. Georges, G. Kotliar, W. Krauth, and M. J. Rozenberg,
Rev. Mod. Phys. {\bf 68}, 13 (1996).

\bibitem{Snoek2011}
M. Snoek, I. Titvinidze, and W. Hofstetter, 
Phys. Rev. B {\bf 83}, 054419 (2011).

\bibitem{Biermann2001}
S. Biermann, A. Georges, A. Lichtenstein, and T. Giamarchi,
Phys. Rev. Lett. {\bf 87}, 276405 (2001).

\bibitem{Orso2007}
G. Orso,
Phys. Rev. Lett. {\bf 98}, 070402 (2007).

\bibitem{Capone2007}
M. Capone, L. de' Medici, and A. Georges,
Phys. Rev. B {\bf 76}, 245116 (2007).

\bibitem{Wolak2010}
M. J. Wolak, V. G. Rousseau, C. Miniatura, B. Gr\'emaud, R. T. Scalettar, and G. G. Batrouni,
Phys. Rev. A {\bf 82}, 013614 (2010).

\bibitem{Kim2009}
D.-H. Kim, P. T\"orm\"a, and J.-P. Martikainen, 
Phys. Rev. Lett. {\bf 102}, 245301 (2009).

\bibitem{Koponen2008}
T. K. Koponen, T. Paananen, J.-P. Martikainen, M. R. Bakhtiari, and P. T\"orm\"a,
New J. Phys. {\bf 10}, 045014 (2008).

\bibitem{Jordens2008}
R. J\"ordens, N. Strohmaier, K. G\"unter, H. Moritz, and T. Esslinger,
Nature (London) {\bf 455}, 204 (2008).

\bibitem{Schneider2008}
U. Schneider, L. Hackerm\"uller, S. Will, Th. Best, I. Bloch, T. A. Costi, R. W. Helmes, D. Rasch, and A. Rosch,
Science {\bf 322}, 1520 (2008).

\bibitem{Edge2009}
J. M. Edge and N. R. Cooper,  
Phys. Rev. Lett. {\bf 103}, 065301 (2009).

\bibitem{Kajala2011}
J. Kajala, F. Massel, and P. T\"orm\"a,
Phys. Rev. A {\bf 84}, 041601(R) (2011).

\end{thebibliography}
\end{document}